\documentstyle[12pt,twoside,fleqn,espcrc1,float,epsfig]{article}

\newcommand{\lsim}{%less than or approx. symbol
 \mathrel{\setbox0=\hbox{$<$}\raise0.6ex\copy0\kern-\wd0
 \lower0.65ex\hbox{$\sim$}}}
\def\be{\begin{equation}}
\def\ee{\end{equation}}
\def\bea{\begin{eqnarray}}
\def\eea{\end{eqnarray}}

\newcommand{\gsim}{%less than or approx. symbol
 \mathrel{\setbox0=\hbox{$>$}\raise0.6ex\copy0\kern-\wd0
 \lower0.65ex\hbox{$\sim$}}}

\newcommand{\AmS}{{\protect\the\textfont2
  A\kern-.1667em\lower.5ex\hbox{M}\kern-.125emS}}

\title{Color Coherent Phenomena with Hadron Beams
\thanks{Dedicated to Koichi Yazaki on the
occasion of his 60th birthday; invited talk presented at the KEK-Tanashi
Symposium on Physics of Hadrons and Nuclei, Tokyo, December 14--17,
1998.}} 
\author{
M.~Strikman\\
{\it Department of Physics, Penn
State University, University Park, PA 16802, USA,\\
and Deutsches Elektronen Synchrotron DESY, Germany\footnote{On leave of
absence from PSU.}}\\
M. Zhalov\\
{\it Institute for Nuclear Physics, St. Petersburg, Russian Federation}}

\begin{document}
\maketitle
\begin{abstract}
We outline major ideas involved in discussion of color coherence 
phenomena (CCP) at intermediate energies. We point out that the recent 
advances in calculating cross sections of hard exclusive processes off light 
nuclei allow to use the lightest nuclei for sensitive tests of CCP.
Consistency of the results of the measurements of color transparency in 
quasielastic A(p,2p) and A(e,e$'$p) processes is emphasized. Evidence for 
presence of significant color fluctuations in nucleons and pions emerging from 
the study of diffractive processes is summarized. 
A new class of hard processes leading to three particle final state 
is suggested for electron and hadron projectiles.
A number of new experiments 
are suggested to probe color fluctuations in hadrons.  
\end{abstract}
%%%%%%%%%%%%%%%%%%%%%%%%%%%%%%%%%%%%%%%%%%%%%%%%%%%%%%%%%%%%%%%%%%%%%%%%%%%%%%%

\section{Introduction}\label{sec:intro}

In this talk we concentrate on the intermediate energy
 Color Coherence  phenomena (CCP) ($E_{inc} \le$ 50 GeV) relevant for
TJNAF, HERMES and KEK hadron facility (JHF) energies.
These studies would be complementary to the studies of CCP at high energies
where a number of such phenomena were recently been observed both in
the scattering of protons - vector meson production at HERA, see the 
recent discussion 
in \cite{Frankfurt}, and in coherent diffraction of a pion to two jets
\cite{Ashery}.

Intermediate energy phenomena are more
 complicated situation - 
dispersion over the size of the produced quark-gluon 
system is likely to be larger,
interaction of the produced system with the target is more complicated.
Also the 
  produced system is not frozen during passage of the nucleus, leading to
dependence of the absorption on the distance from the interaction point.
One can treat this as a mere complication - but 
in
fact this is 
a separate aspect of QCD which we miss in high-energy processes and which 
 deserves dedicated studies.

\section{Color transparency phenomena}

To describe cross sections of hard processes in QCD  one has to introduce
the Fock space decomposition of the light-cone hadron wave function
over configurations containing different number of constituents.
Probably the most interesting components in hadrons
are those which contain
minimal number of constituents. They   determine asymptotic
behavior of various
exclusive hard processes  such as  electromagnetic form factors.
One can expect that at very large momentum transfers point-like
(small size components) (PLC) of the hadron wave function should dominate in 
the scattering. To check this assumption it was suggested by 
Brodsky \cite{Brodsky82} and Mueller 
\cite{Mueller82} to study quasi-exclusive hard reactions
$l(h) +A \rightarrow l(h) + p +(A-1)^*$. If the energies and momentum 
transfers are large enough one expects that projectile and ejected nucleon
travel through the nucleus in  
point-like (small size) configurations, resulting in a cross section
 proportional to $A$.

In accessing the range of applicability of this 
approximation one has to address two questions: (i) Can PLC be treated
 as a frozen during the passage of the nucleus, (ii) At what 
momentum transfer
PLC's dominate in the elementary amplitude.

\subsection{Expansion effect}
The current color transparency experiments are performed in 
the kinematics where expansion of the produced
small system is very important (essential longitudinal distances are not
 large enough) and strongly suppresses color transparency effect
\cite{FLFS88,jm90}. 

The maximal longitudinal  distance for  which coherence effects are still 
present is determined by the minimal characteristic internal excitation
energies of the hadron h. The estimates \cite{FLFS88,jm90} show that 
for the case of a nucleon
ejectile
  coherence is completely lost at the distances
$l_c \sim (0.3\div0.5)\cdot p_h$ fm,
where $p_h$ is measured in GeV/c.

To describe the effect of the loss of coherence two complementary 
languages were suggested. In Ref.~\cite{FLFS88} based on 
the quark-gluon representation of PLC wave function it was argued that the
main effect is quantum diffusion of the wave packet so that  
\bea
\sigma^{PLC}(Z) =(\sigma_{hard} + {Z\over l_c}[\sigma 
-\sigma_{hard}])
\theta(l_c- Z) +\sigma\theta\left(Z-l_c\right). \label{eq:sigdif}
\label{eq1}
\eea
This equation is justified for hard stage of time development
in the leading logarithmic
approximation when perturbative QCD can
be applied \cite{FLFS88,BM88,FS88,DKMT}.
One can expect that Eq.(\ref{eq:sigdif})
smoothly  interpolates between the hard and soft regimes.
A sudden change of $\sigma^{PLC}$ would be inconsistent with
the observation of an early (relatively low Q$^2$)
Bjorken scaling \cite{FS88}. Eq.(\ref{eq1}) implicitly incorporates the
 geometric scaling for the PLC-nucleon interactions which
for the discussed energy range include nonperturbative
effects. However the discussed approximation for the
expansion effects is  oversimplified, see discussion in section \ref{2.3}.

The time development of the $PLC$ can also be 
obtained by modeling the ejectile-nucleus interaction using  
 a baryonic basis for the wave function of PLC:
\begin{eqnarray}
\left| \Psi_{PLC}(t)\right>=\Sigma_{i=1}^{\infty} a_i 
\exp(iE_it)\left| \Psi_{i} \right>=
 \exp(iE_1t)\Sigma_{i=1}^{\infty} a_i 
\exp\left({i(m_i^2-m_1^2)t\over 2P}\right)\left| \Psi_{i} \right>,
\end{eqnarray}
where $\left| \Psi_{i} \right>$ are the Hamiltonian eigenstates  with
masses $m_i$, and $P$ is the momentum of PLC which satisfies 
$P \gg m_i$.  As soon as the relative phases of 
the different hadronic components
become large (of the order of one) the coherence is likely to be lost.
It was however suggested  by B.Pire and J.Ralston
 that coherence may be sustained over much larger distances, see contribution
of B.Pire \cite{pire} and references therein. One rather special
example when coherence 
is sustained indefinitely is the harmonic oscillator - in this case 
coherence is sustained due to the  equidistance of the energy levels.

Color transparency requires that the amplitude 
of $PLC- N $ interaction is small, leading to a number
of constraints on the amplitudes of 
 $i+N \to j+N$ scattering
\cite{jm90,FS91,GFMS92,boffi,jm94}. If many intermediate states are
 included in the
hadron-basis model,
 the numerical results of two approaches for $\sigma_{eff}$
turn out to be quite similar. 

It is worth emphasizing that though
both approaches model 
certain aspects of dynamics of expansion, a 
complete treatment of this phenomenon in QCD is  missing so far.
In particular, the 
 phenomenon of spontaneously broken chiral symmetry may
 lead to presence of two scales in the rate of expansion, one 
corresponding to regime where
quarks can be treated as massless, and another 
where virtualities become small enough and quarks 
acquire effective masses of the order of 300 MeV.

\subsection{Are small configurations been made?}

Current pQCD
 analyses indicate that the leading twist approximation for
the pion  form factor could become applicable (optimistically)  at 
$Q^2 \ge 10-20 GeV^2$, for the recent analysis see Ref.~ \cite{JRS}.
For the nucleon case even larger $Q^2$ are likely to be necessary.
However
this does not preclude PLC from being
 relevant for smaller $Q^2$. In fact, in a wide range of models
of a nucleon, such as constituent quark models with singular 
(gluon exchange type) short-range interaction, pion cloud models, the 
configurations of  sizes substantially smaller than the average
 one dominate in the form factor at $Q^2 \ge 3-4 GeV^2$, see discussion
in Ref.~ \cite{fms}.
Message from the QCD sum rule model 
calculations of the nucleon form factor is
more ambiguous.

Theoretical expectations for the large
 angle hadron-hadron scattering are  even less clear. Irregularities in the
 energy dependence of the $pp$ scattering for $\theta_{c.m.}=90^o$,
and large spin effects have 
lead to suggestion of presence of two interfering mechanisms
in this process \cite{rp88,bdet} corresponding to interaction of nucleon in
configurations of small and large sizes.

\subsection{Cross section of PLC-nucleon interaction}
\label{2.3} 
The cross section of high-energy 
PLC-nucleon interaction can be expressed in pQCD
through the gluon density in the nucleon. In particular for
the interaction of the $q\bar q$ pair of the transverse size $b=
r_{q}^{t}- r_{\bar q}^t$ \cite{BBFS93}
\begin{equation}
\sigma_{q\bar q,N}(E_{inc})={\pi^2\over
 3}b^2\alpha_s(Q^2)xG_N(x,Q^2\equiv{\lambda\over b^2}),
\end{equation}
where $\lambda(x \approx 10^{-3})\approx  9,
x={Q^2\over 2m_NE_{inc}}$.
At intermediate energies the gluon density
enters at rather large $x$, leading to a further suppression of this
cross section as compared to higher energies.
All this indicates that the geometrical scaling - $\sigma \propto b^2$ - 
could be violated at intermediate energies leading to a
lack of smooth connection between
cross section of interaction of hadrons in average and in PL configurations.
Emission of gluons in the process of expansion, and hence interaction
of the wave package in configurations containing extra gluons
could become important.
Also  the
 dominance of the two gluon coupling to a PLC  is not obvious for such 
energies -
a  competing mechanism of interaction of PLC
could be emission of a pion in PLC 
which contains similar suppression factor as the
the coupling to PLC via two gluons
\cite{FS85}. 
Furthermore at the last stage of the expansion and formation of
 the outgoing hadron the spontaneous chiral symmetry breaking
can affect both the rate of expansion and the interaction of the
 wave packet with the target.
Overall one can expect that interaction of PLC with nucleons
at these energies is 
 more complicated than, for example, in the case of the high-energy 
vector meson production by longitudinally polarized photons.

\subsection{Experimental data}
The current data on 
 $A(p,2p)$ reaction \cite{Car88} seem to support increase of transparency  
at $p_{inc}=6,10 GeV/c$ as compared to that observed at $T_p=1\, GeV$, 
see discussion  in \cite{FSZ94}. 
The magnitude of the effect can be  easily
 described in the color transparency
models which include the expansion effect.
Description of the 12 GeV/c data where a drop of  the
 transparency is indicated requires invoking interplay of contributions
of PLC and large size configurations as suggested in \cite{rp88,bdet}.
Such a description was achieved in Ref.\cite{jm94}. The first
data of the EVA experiment at BNL\cite{EVA98} confirmed 
a significantly larger value of transparency than the one
expected in the Glauber 
approximation. At the same time the data indicate 
a strong variation of the transparency as a function of the
center of mass scattering angle in the
studied angular range of $90^o\ge \theta_{c.m.}\ge 84^o$
for $p_{inc}=5.9 GeV/c$. Such a variation poses a challenge to all models since
the elementary $pp$ reaction does not show any variation
 of the energy dependence in this angular range.

First electron $A(e,e'p)$ experiment, NE-18,
 aimed at looking for color transparency was performed at 
SLAC ~\cite{NE18}. Maximum $Q^2$ in this experiment is $\approx 7 GeV^2$
which corresponds to $l_{coh} \le 2 fm$. 
In
this kinematics 
the color transparency models which included expansion effects 
predicted a  rather small increase of the transparency, 
see for example \cite{FS88}.
This prediction is consistent  with the  NE-18
data. However these data are
 not sufficiently accurate  either to confirm or to rule out color 
transparency on the level predicted by the realistic color 
transparency models, for the 
detailed discussion see review \cite{fms}. 
It should be pointed out that 
 the energy resolution of this experiment was insufficient
 to separate the levels, which further complicates interpretation
 of this experiment since the theoretical uncertainties
are larger in this case.
Also the restricted recoil energy   range of the experiment leads to
potential complications due to the nuclear quenching effects
 see discussion in \cite{nikhef}.

\section{High-momentum transfer exclusive processes at JHF }

i) Fixed $t$ color transparency studies.

At high energies cross section of the elastic $pp$ scattering at fixed
$\theta_{c.m.}$ drops rapidly with s ($d\sigma/dt\propto s^{-10}$). As a result
measurements  at $\theta_{c.m.} \sim 90^o$ would be hardly
possible. At the same time it is expected that
PLC will dominate in the interaction already at $-t \ge -t_0 \sim $ few
GeV$^2$.
This corresponds to the kinematics when both the projectile and the
leading scattered proton are very fast and can be considered as nearly
frozen 
in the interaction process. Only the recoiling nucleon is rather
strongly absorbed. Its absorption should be similar to the one 
in the $A(e,e'p)$ reaction at the corresponding $Q^2$. Hence one
expects that nuclear transparency should strongly increase as a function of
incident energy for fixed $t$, see Fig.1.

\begin{figure}
\vspace{-1.1cm}
    \begin{center} 
        \leavevmode 
        \epsfxsize=0.65\hsize 
        \epsfbox{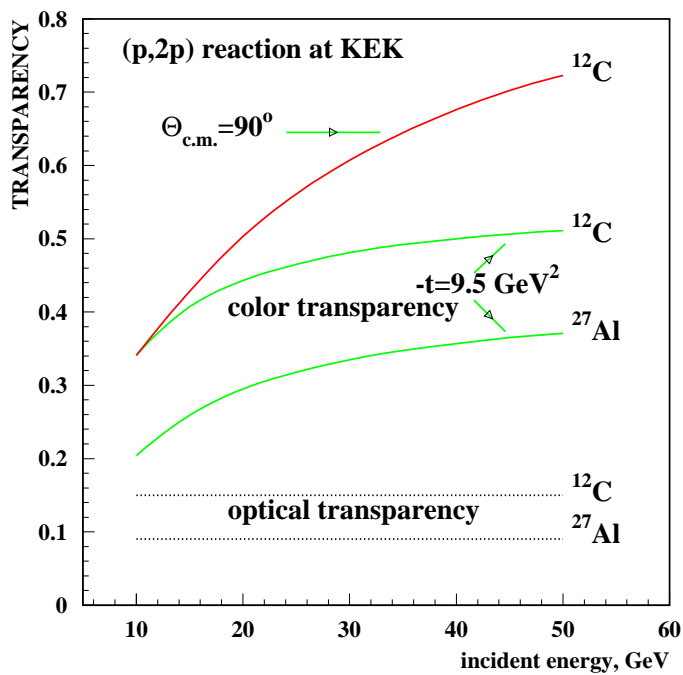}
          \epsfxsize=0.65\hsize
           \epsfbox{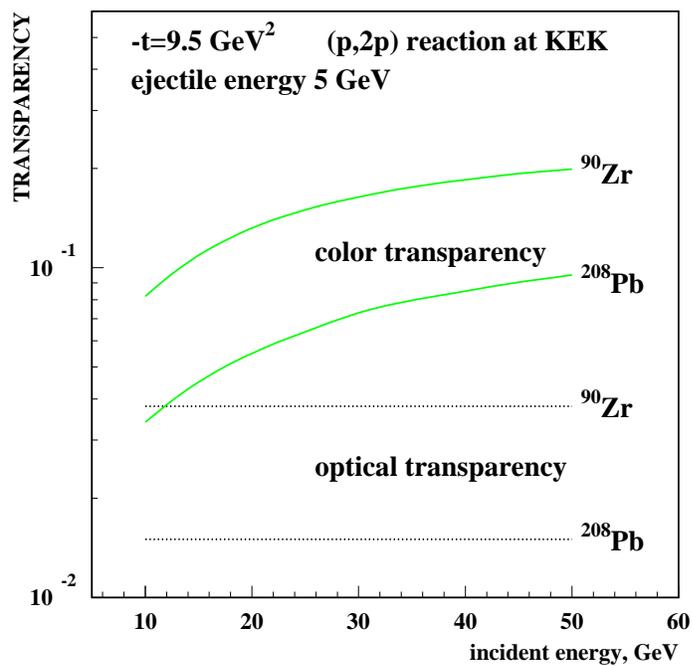}
    \end{center} 
\caption{ {\em Energy dependence of the nuclear transparency
calculated in the quantum diffusion model with $\Delta m^2=0.7 GeV^2$
as compared to the expectations of the Glauber model.}}
\end{figure}
Such studies would require a detector with momentum resolution 
$\Delta p/p\sim 1\%$ and angular acceptance of $5^o \le
\theta_{lab}\le 30^o$.

ii) Use of different projectiles is necessary, since different hadrons
have different PLC probabilities. Furthermore, the relative
importance of various
reaction
mechanisms may  vary with the projectile. In particular,
experimental studies of two-body large angle reactions
\cite{Car2} have demonstrated that reactions where quark exchange is
allowed have substantially larger cross sections than reactions where 
quark exchanges are forbidden.
 The diagrams corresponding to the quark counting rules are expected 
to be dominated by small interquark distances,  
while the dominance of the 
 small interquark distances arises in the diagrams with
multiple  gluon exchanges due to 
Sudakov form factor effects only. Thus the 
energy
 dependence of the nuclear transparency 
in different channels may be quite different. Consequently,  
 the  experimental study of the
reactions like   $p+A\rightarrow \Delta^{o} +p + (A-1)^*$  would be
extremely
 important.

It would be also very interesting to study nuclear transparency with
kaon, and antiproton projectiles. In particular, this may help to 
understand why the elastic $\bar p p$ cross section is suppressed as
compared to the elastic $pp$ cross section at least by a factor of 100
at $p_{inc}= 6$ GeV/c.

\noindent 
iii) One important consequence of the CT picture is that cross sections
for production of 
excited
states in high-energy wide
angle reactions
\cite{jm90,FS91} are large.
So it is important to look for the processes like
$p+A\rightarrow N^{*} (N\pi, N\pi\pi) +N + (A-1)^*$. Knowledge of the
relative abundance of
resonances and continuum would help in building more realistic models for the
expansion of PLC.

\section{Probing the PLC in the 
high energy scattering from the lightest nuclei}

To investigate the implication of the QCD physics for nuclear reactions 
at not extremely high energies one has to find a way to
fight effects of space-time evolution of the 
the quark-gluon system
 produced in a hard $\gamma^*(h)N$
 scattering. As we discussed above 
this evolution results in an expansion of a PLC (assuming that it was 
produced in the hard interaction) 
to an average size configuration before it could reach 
another nucleon. In this case one would not be able to observe 
any color coherence/transparency effects.

Hence an optimal strategy seems to be
to select the processes where propagating system may interact with a residual
system close to the primary interaction point. This minimizes 
expansion effects and 
therefore  allows 
to investigate the PLC at an earlier stage of formation of the final state.
However, theoretical methods which were successful in  the medium-energy   
nuclear physics should be upgraded in order to describe processes  where  
energies transferred to a nuclear target are $\ge  few$ $GeV$. 

Hence a new theoretical approach for calculation of the 
high energy hard semi-exclusive reactions off the lightest nuclei
has recently been developed 
\cite{FGMSS95,ggraphs}. 
In the nucleus lab. frame these reactions correspond to the kinematics where
the momentum transfered to the nucleus $q=p_{inc}-p_{fin.proj}$ 
and the momentum of one 
of the hadrons in the final state $p_f$ (the fast hadron) are:
$q,p_f\sim~few~$GeV/c and ($q\approx p_f$) while 
the excitation 
energy of the residual nuclear system $E_r$ is $\sim~hundreds~$MeV: 
\begin{equation}
{q_-\over q_+} \ \ , \ \ {p_{f-}\over p_{f+}} \ll 1.
\label{leap}
\end{equation}
Here
$k_{\pm} = k_0 \pm k_z$ ($k=q,p_f$) and $z$ is in the
$\vec q$ direction.  
The condition  (\ref{leap}) provides
the small parameter for 
our calculations. Using this condition it is straightforward to 
demonstrate that within the accuracy 
$O({q_-\over q_+},{p_{f-}\over p_{f+}})$ 
the sum of  all possible Feynman diagrams which 
describe the particular hard semi-exclusive reaction off nuclei 
could be reduced \cite{Gribov,Bertocchi,ggraphs} to the sum of the restricted 
number of diagrams where the soft rescattering of the fast hadron (h) with 
slow nucleons of the target can be described by the invariant amplitude 
of  the elastic $hN$ scattering, $f_{hN}$.
Thus one ends up with a set of  covariant diagrams for  rescatterings
where usual Feynman rules are  applied and vertices $f_i$ correspond to 
above mentioned phenomenological amplitudes.

The diagrams of this type has  two crucial
features which reflect
 the high-energy nature of the scattering process and  the "soft" nature 
of rescattering of fast hadrons with the slow nucleons of the target:

 $\bullet$
 For each vertex
 $f_i$, the $"-"$ component of the scattered 
particle's  momentum is  conserved
 up to the factor $O({q_-\over q_+},{p_{f-}\over p_{f+}})$.
Indeed the energy-momentum conservation for the $f_k$ vertices implies
that the 
$"-"$ components of the rescattered slow $k$ nucleons are proportional  to 
${p_{f-}\over p_{f+}}\approx {m_f^2\over p_{f+}^2}$.

$\bullet$
The   propagator of the fast hadron ( knocked-out    
nucleon - $D(p_1+q)^{-1}$) decouples from the slow (nuclear) 
part of the reaction. Such a decoupling allows to  use the closure
over the  intermediate nuclear states, and thus 
to reduce the covariant nuclear vertices to the nuclear wave functions.
As a result of relativistic kinematics we
we find
that 
at fixed $-t/s$ and high energy transfer limit, amplitude
depends only on the  $p_{1-}$ component of the struck nucleon 
momentum. 
Therefore at fixed $p_{-}$ we found  effective factorization of  
the high-energy propagator from low energy intermediate nuclear part 
whose excitation energy on light cone is defined  by the 
$p_{1+}$ \cite{FS81,FS88}.  Such a decoupling is valid
 for any 
values of the target nucleon Fermi momenta 
and not restricted to the nonrelativistic limit
${p_1^2\over 2m^2}\ll 1$.

These are two of the major differences 
of the Feynman diagram approach from  the conventional Glauber 
approximation \cite{Glauber}, which is  applicable only for the 
case when the Fermi motion of the spectator nucleons is neglected.

Two applications of the Feynman
diagram formalism are
(i) the 
knock-out $(e,e'p)$ reaction from the lightest nuclei,  and  (ii)
large angle $^2H(p,2p)$ reaction \cite{pd}.

High $t$ knock-out reaction from the deuteron is described by 
eight diagrams: the  plane wave impulse approximation (IA) diagram $F_A$ 
(where knocked-out fast nucleon does not 
interact with  spectator nucleon), and by rescattering diagrams $F_B$ (Fig.2)
where projectile scatters off both nucleons or
the final state rescatterings of the scattered nucleon off the would
be spectator occur. The diagrams
 with single soft interaction lead to the screening
of the impulse approximation diagram, while the diagrams
 with two soft interaction
have the same sign as the impulse approximation diagram.
\begin{figure}[p]
\centerline{
\epsfig{file=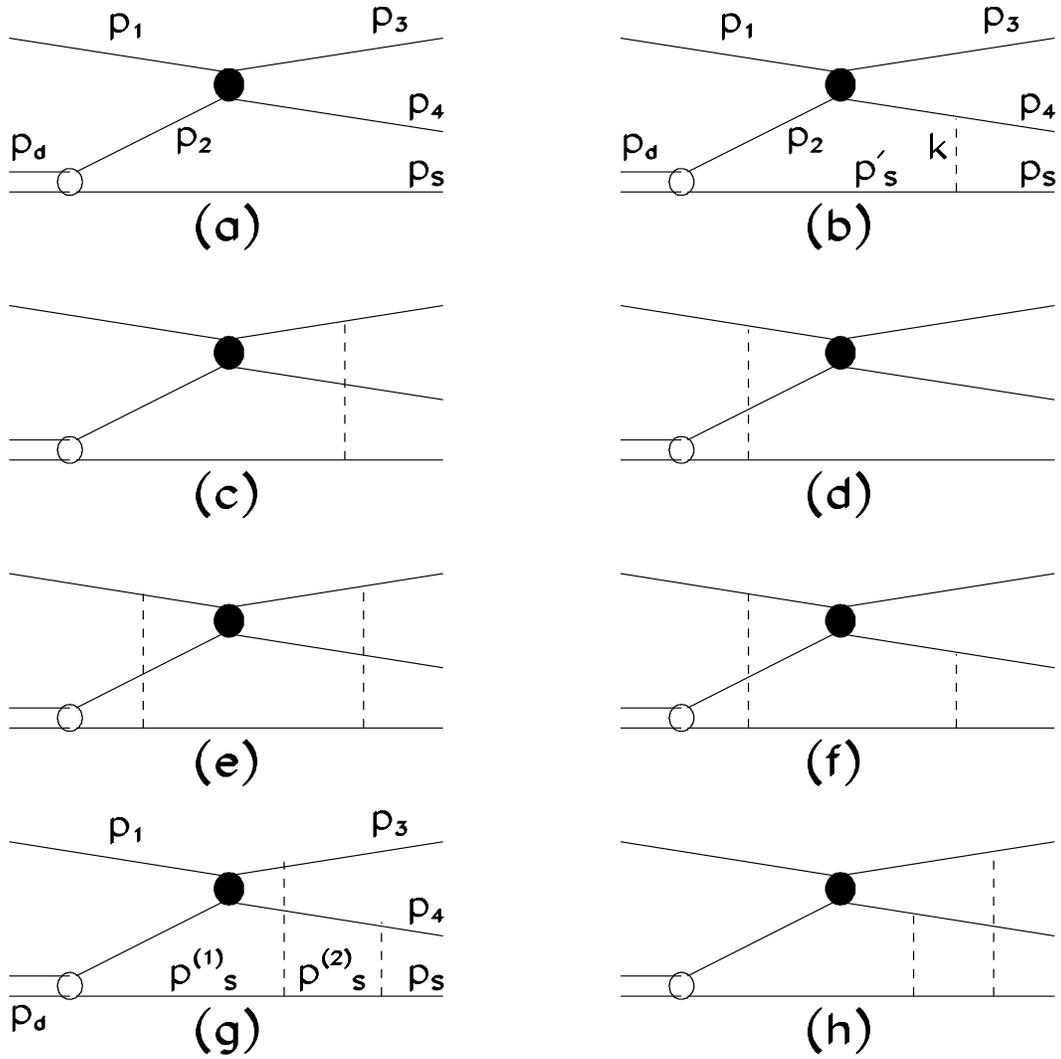,width=14cm,height=14cm}
}
\caption{ {\em Feynman diagrams of the  eikonal approximation for $^2H(p,pp)n$ 
scattering. The dashed lines describe the amplitude of $NN$ scattering, 
the full  circles represent the hard $pp$ scattering  amplitude.}} 
\end{figure}
Applying the Feynman rules for the
rescattering diagram and switching to the 
coordinate space representation we obtain an
expression which resembles familiar expression from  the Glauber
approach
(for simplicity we give expression for the $^2H(e,e'pn)$ process where
only one rescattering diagram is present:
\begin{eqnarray}
F^{(B)} =  -  \int  d^3x 
\phi_d(x_)F^{em}_1(Q^2)\Theta(z)\Gamma_{NN}(x,\Delta). 
\end{eqnarray}
The key distinction  is emergence of the 
modified profile function $\Gamma_{NN}(x,\Delta)$ which 
differs from the conventional Glauber profile function as 
follows
\begin{equation}
\Gamma_{NN}(x,\Delta) = e^{i\Delta z}  \Gamma^{(GA)}_{NN}(x).
\label{mdi}
\end{equation}
Here  $\Delta$ is given by  defined according to
\begin{equation}
\Delta = (E_s-m){E_f\over p_{fz}}
- (p_{st}-p'_{st}){p_{ft}\over p_{fz}}.
\label{eq.6}
\end{equation} 
and describes the 
excitation of the residual nuclear system 
(spectator nucleon in  the 
case of the deuteron target). Here subscripts $s$ and $f$ 
refer to the spectator system and the knocked out nucleon.
Due to
 the steep momentum dependence of the deuteron wave function, the
factor $e^{i\Delta z}$ in 
Eq.(\ref{mdi}) results in a significant difference 
 between 
the predictions of the conventional Glauber approach and the present approach 
based on the Feynman diagram technique, see comparison
in Ref. \cite{FGMSS95}.

One finds that the $^2H(p,pn)$ reaction is very sensitive to the
effects of color transparency and can provide a detailed mapping of
the pattern of the expansion. The expected effects are large, see
e.g. Fig.3.
\begin{figure}
\centerline{
\epsfig{file=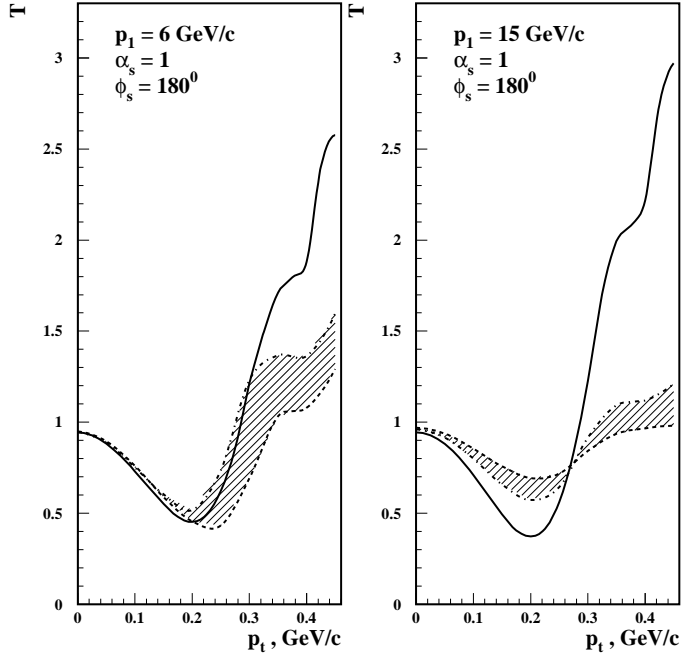,width=10cm,height=10cm}
}
\caption{ {\em The $p_t$ dependence of $T$ at $\alpha_s=1$.
The solid line is for the  elastic eikonal approximation 
which neglects color transparency 
effects. The shaded area corresponds to $T$ 
calculated within  the quantum diffusion model of CT.
Dashed and dash-dotted curves correspond to QDM calculation 
with $\Delta M^2=0.7$ and   $\Delta M^2=1.1~GeV^2$ respectively.
}}
\end{figure}
Strong effects are also predicted for the out of plane kinematics.

Overall we now have  a reliable formalism for  the treating 
final interaction of the 
knock out nucleons in electron and hadron nucleus reactions. Hence
a set of experiments with lightest and medium size nuclei
could allow
to perform a detail mapping of the space-time 
evolution of the wave packet produced in the elementary hard process.

\section{Color Fluctuations in Hadrons and Diffraction}

Since coherent length in high-energy processes is large,
a fast hadron
 can be considered as
a superposition of constituents
 frozen during such collisions. Therefore
 existence of parton configurations within  hadrons having
 small 
interaction cross section (as confirmed by the diffractive
electroproduction of
 $\rho-$meson 
production) implies that  significant fluctuations should be present in the 
intensity of interaction of fast hadrons with targets. It is convenient to
introduce a bulk   characteristic of these fluctuations - the 
probability for a hadron to interact with 
certain intensity, $P(\sigma)$, see review in \cite{fms}.
 One can describe $P(\sigma)$ in terms of its moments: $\langle
\sigma^n\rangle =\int \sigma^n P(\sigma) d\sigma $. The zeroth moment is unity,
by conservation of probability, and the first corresponds to the total
hadron-nucleon cross section $\sigma_{tot}$. The second and third
moments has been
determined from available diffractive dissociation data for scattering off
protons and deuterons as well as from the measurements of inelastic shadowing
for $\sigma_{tot}(hD)$.
 Different
determinations \cite{BBFHS93}
 give consistent values for the variance of the distribution: $\omega_{\sigma} \equiv
\left( \langle \sigma^2 \rangle -\langle \sigma \rangle^2 \right) /  \langle
\sigma^2\rangle$ , with $\omega_{\sigma}(p) \sim 0.25$ and
$\omega_{\sigma}(\pi) \sim 0.4$, near 200 GeV/c momenta. 
The behavior of $P(\sigma)$ for $\sigma \rightarrow 0$
is determined by the interaction with minimal Fock configurations
 in the hadrons, leading to $P(\sigma) \propto \sigma^{n-2}$ where $n$
 is number of constituents in the hadron. $P(\sigma)$ estimated from
 data is broad; in line with the view that
different size configurations interact with widely varying cross
sections.  Several analyses seem to confirm this picture.

$\bullet $ For
the
 pion projectile it is possible to calculate $P(\sigma)$ for small $\sigma$
along the same 
lines as for the $\rho-$meson production \cite{BBFS93}, 
\begin{eqnarray}
P(\sigma \ll \left<\sigma\right>)
=\frac{6f^2_{\pi}}{5 \alpha_s(4k^{2}_{t})\bar x
  G_{N}(\bar x, 4k^{2}_{t})},
\label{p0}
\end{eqnarray}
where $\alpha_s(4k^{2}_{t})$ is the QCD running coupling constant, and
$G_N(\bar x,4k^{2}_{t})$ is the gluon distribution in the nucleon;
 $\bar x= 4k^2_t/s_{\pi N}; k^2_t \approx 1/b^2$ where $f_{\pi}$ is
 the constant for $\pi \rightarrow
\mu \nu$ decay. This  result is in a reasonable agreement with
determination from diffractive data.

\hskip 1 cm
Nuclear inelastic coherent diffractive hadron production 
provides another 
 nontrivial experimental test of the concept of color fluctuations.

$\bullet $ 
The total diffractive cross section can  be computed using 
$P(\sigma)$ 
\cite{FMS93b}. Results of the calculations agree reasonably with available 
data
for the A-dependence of the semi-inclusive production
 for pion and proton projectiles.

$\bullet$ Coherent 
diffractive cross section emulsion data \cite{boos} for 400 GeV protons 
on $\left<A\right> \approx$ 50
 nuclei give large cross sections consistent with the
color fluctuation expectations \cite{FMS93b}. 

$\bullet $  It was shown in Ref.~\cite{sg} that 
the color fluctuation cross section expectations 
are in good agreement with the forward angle 
cross section data for $p +^4He \rightarrow
X + ^4He$ data \cite {buj}. 

Obviously 
 much more detailed experimental studies are necessary to check 
the details of the picture.

\section{Star dust processes}
 So far the studies of hard
 exclusive processes were concentrated on two-body reactions.
It is natural to move one step further and 
ask a question whether collapsing of 3 valence quarks 
to a small size configuration in a nucleon or 
of valence 
 $q \bar q$ in a meson would result in disappearance of other constituents?
Such scenario would be natural in quantum electrodynamics 
for the case of positronium - the photon field disappears in the case when 
electron and positron are close together.
However  in QCD where interactions at large distances are strong
it is possible that non-minimal Fock components of the 
hadron contain configurations with small color singlet  clusters.
To investigate this question we suggest  a study of a new 
class of hard exclusive processes where large momentum is transfered to 
subsystem of the hadron and residual system has a finite mass 
not increasing with hardness of the process \cite{baryon95}.  
Examples of such reactions include $e +p \rightarrow e + M +B$,
$h +p \rightarrow h' + M +B$ where M is a meson carrying
most of the transfered momentum, while $B$ is recoil system which
 mass and momentum are kept fixed in the target rest frame 
(or in the projectile rest frame),see for example Fig.4.

\begin{figure}
    \begin{center}
        \leavevmode
        \epsfxsize=.5\hsize
        \epsfbox{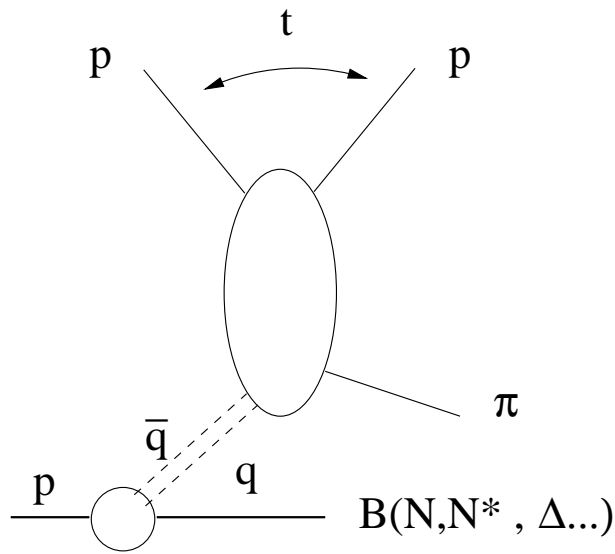}
    \end{center}
\caption{\it Production of fast pion and recoiling baryonic system.}
\end{figure}

The discussed limit is
\begin{equation}
\alpha_B={E_B-p_{3B}\over m_N}=const, {s_{N_{fin}M}\over s}=1-\alpha_B,
s\to \infty, -t/s=const
\end{equation}

Reverse processes are also possible when a fast baryon is ejected from
the target: $e +p \rightarrow e + B +M$,
$h +p \rightarrow h' + B +M$ - Fig.5.
\begin{figure}
    \begin{center}
        \leavevmode
        \epsfxsize=.6\hsize
        \epsfbox{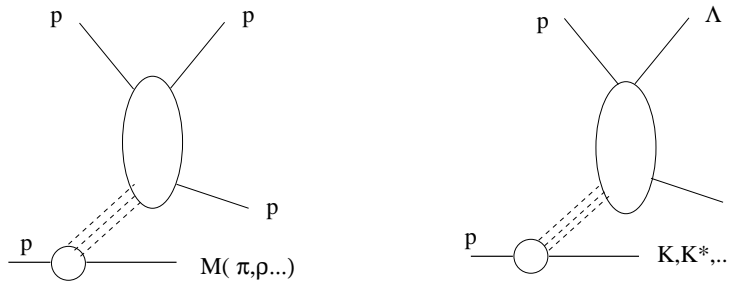}
    \end{center}
\caption{\it Production of fast baryon and recoiling mesonic system.}
\end{figure}

Processes of both kinds emerge naturally in the pion cloud 
picture
of the nucleon (a baryon)
as scattering of the nucleon core with a pion been a spectator or visa
 versa
scattering of a pion with a nucleon 
($\Delta$-isobar) spectator. Within the pion cloud model
cross section of these reactions 
should
have cross section comparable to
 the cross section of two body large angle process.

Generally one could expect comparable cross sections for production in a recoil
of a pion and heavier states. Investigation of these reactions
is important also for study of $(e,e'p), (p,2p)$ reactions
where they produce a background for high excitation energies 
$\ge (200\div 300) MeV$.

Recently the QCD factorization theorem was proven for the DIS
exclusive processes initiated by the 
$\gamma_L +T\to M +B$ \cite{CFS}. The cross section is expressed through the 
convolution of the $q\bar q$ component of the meson wave function,
hard blob, and the skewed parton densities in the nucleon.
The quark skewed parton densities are defined as:
\begin{eqnarray}
   f_{ij/p}(x_{1},x_{2},t,\mu )  &=&
   \int _{-\infty }^{\infty } \frac {dy^{-}}{4\pi }
   \;
   e^{-ix_{2}p^{+}y^{-}}
   \langle p'|\; T {\bar \psi }(0,y^{-},{\bf 0}_{T})\gamma ^{+}
     {\cal P} \psi (0)\; |p\rangle  ,
\label{pdf.q.def}
\end{eqnarray}
where $\cal P$ is a path-ordered exponential of the gluon field
along the light-like line joining the two operators for the quarks
of flavors $i,j$ and the final state could be any baryon allowed by the
quantum numbers. $x_{1},x_{2}$ are the light-cone fractions of the
quark and antiquark. For the process under discussion $x_i$ satisfy
\begin{equation}
x_1-x_2=1-\alpha_B.
\end{equation}

Recently calculations of some of these densities were performed in the
QCD chiral model \cite{Bochum}. They indicate that probabilities to
find a $q\bar q$ pair in the nucleon in a PLC configuration is
comparable to the expectations of the pion model in a wide range of 
$\alpha_B$.

If the corresponding two-body  process is dominated by the
the scattering in PLC we may expect a scaling relation between
the cross sections of the processes induced by longitudinally
polarized photons and hadrons:
\begin{equation}
{{d\sigma^{pp\to p+\pi +B}\over d \alpha_Bd^2p_{tB}d\theta_{c.m.}(p\pi)}
\over {d\sigma^{p\pi\to p+\pi}\over d\theta_{c.m.}}(s_{p\pi})}
={{d\sigma^{\gamma^*_L+p \to \pi +B}(Q^2)\over d\alpha_Bd^2p_t}
\over \sigma^{\gamma^*_L+\pi \to \pi}(Q^2)}.
\end{equation}
Similar scaling relations are expected for various hadronic
projectiles.

Studies of the ``star dust'' processes provide a unique way to study
parton color singlet correlations in nuclei. To be able to reach the
scaling region incident energies of the order 20 GeV are necessary 
to insure that the energy in the hard blob is large enough.

To summarize,
a diverse program of studies of dynamics of hard coherent processes
is possible at JHF energies. One would be able to get a better insight
into the properties of point-like configurations in the hadrons,
investigate the space time evolution of small quark-gluon
wave packages, and also to start investigation of small color singlet
clusters in hadrons.
\section*{Acknowledgments}

We would like to thank L.Frankfurt, G.Miller and M.Sargsian for numerous 
discussions. This work is supported by the US Department of Energy under 
contracts DE-FG02-93ER40771. We would like also thank DESY for
hospitality during time this manuscript was prepared.

\end{document}